\documentclass[12pt]{article}
\usepackage{graphicx,amssymb,amsfonts,amsmath,epsfig,color}

\makeatletter
\DeclareSymbolFont{AMSb}{U}{msb}{m}{n}
\DeclareSymbolFontAlphabet{\mathbb}{AMSb}
\renewcommand{\section}{\@startsection{section}{1}{\z@}%
                                    {-7ex \@plus -1ex \@minus -.2ex}%
                                    {2.5ex \@plus.2ex}%
                                    {\normalfont\large\scshape\centering}}
\renewcommand{\subsection}{\@startsection{subsection}{2}{\z@}%
                                       {-5ex \@plus -1ex \@minus -.2ex}%
                                       {1.5ex \@plus.2ex}%
                                       {\normalfont\normalsize\scshape}}
\renewcommand{\subsubsection}{\@startsection{subsubsection}{3}{\z@}%
                                       {-5ex \@plus -1ex \@minus -.2ex}%
                                       {1.5ex \@plus.2ex}%
                                       {\normalfont\normalsize\scshape}}

\renewcommand\@seccntformat[1]{\ignorespaces\csname #1name\endcsname\space
                               \csname the#1\endcsname.\quad}   
\newdimen\captionmargin
\newdimen\captionindent
\newdimen\captionwidth
\newcommand{\captionfont}{\slshape}
\newcommand\@captionlabel[1]{\textsc{#1:}\space}
\long\def\@makecaption#1#2{%
  \vskip\abovecaptionskip
  \captionwidth\hsize
  \advance\captionwidth -2\captionmargin
  \sbox\@tempboxa{\@captionlabel{#1}\captionfont #2}%
  \ifdim \wd\@tempboxa >\captionwidth
    \ifdim\captionindent>\z@
      \advance\captionwidth -\captionindent
      \hskip\captionindent
    \fi
    \hskip\captionmargin
    \parbox[t]{\captionwidth}{\leavevmode\hskip-\captionindent
      \@captionlabel{#1}\captionfont #2}%
  \else
    \global \@minipagefalse
    \hb@xt@\hsize{\hfil\box\@tempboxa\hfil}%
  \fi
  \vskip\belowcaptionskip}
\def\eqnarray{%
   \stepcounter{equation}%
   \def\@currentlabel{\p@equation\theequation}%
   \global\@eqnswtrue
   \m@th
   \global\@eqcnt\z@
   \tabskip\@centering
   \let\\\@eqncr
   $$\everycr{}\halign to\displaywidth\bgroup
       \hskip\@centering$\displaystyle\tabskip\z@skip{##}$\@eqnsel
      &\global\@eqcnt\@ne$\;\hfil{##}$\hfil
      &\global\@eqcnt\tw@$\;\displaystyle{##}$\hfil\tabskip\@centering
      &\global\@eqcnt\thr@@ \hb@xt@\z@\bgroup\hss##\egroup
         \tabskip\z@skip
      \cr}
\ifcase \@ptsize
\or
\or
\fi
  \divide\@tempdima\baselineskip
  \@tempcnta=\@tempdima
\makeatother
\begin{document}

\renewcommand{\theequation}{\arabic{section}.\arabic{equation}}
\renewcommand{\thefigure}{\arabic{figure}}
\newcommand{\gapprox}{%
\mathrel{%
\setbox0=\hbox{$>$}\raise0.6ex\copy0\kern-\wd0\lower0.65ex\hbox{$\sim$}}}
\textwidth 165mm \textheight 220mm \topmargin 0pt \oddsidemargin 2mm
\def\ib{{\bar \imath}}
\def\jb{{\bar \jmath}}

\newcommand{\ft}[2]{{\textstyle\frac{#1}{#2}}}
\newcommand{\be}{\begin{equation}}
\newcommand{\ee}{\end{equation}}
\newcommand{\bea}{\begin{eqnarray}}
\newcommand{\eea}{\end{eqnarray}}
\newcommand{\Identity}{{1\!\rm l}}
\newcommand{\cx}{\overset{\circ}{x}_2}
\def\CN{$\mathcal{N}$}
\def\CH{$\mathcal{H}$}
\def\hg{\hat{g}}
\newcommand{\bref}[1]{(\ref{#1})}
\def\espai{\;\;\;\;\;\;}
\def\zespai{\;\;\;\;}
\def\avall{\vspace{0.5cm}}
\newtheorem{theorem}{Theorem}
\newtheorem{acknowledgement}{Acknowledgment}
\newtheorem{algorithm}{Algorithm}
\newtheorem{axiom}{Axiom}
\newtheorem{case}{Case}
\newtheorem{claim}{Claim}
\newtheorem{conclusion}{Conclusion}
\newtheorem{condition}{Condition}
\newtheorem{conjecture}{Conjecture}
\newtheorem{coro}{Corollary}
\newtheorem{criterion}{Criterion}
\newtheorem{defi}{Definition}
\newtheorem{example}{Example}
\newtheorem{exercise}{Exercise}
\newtheorem{lemma}{Lemma}
\newtheorem{notation}{Notation}
\newtheorem{problem}{Problem}
\newtheorem{propos}{Proposition}
\newtheorem{rem}{{\it Remark}}
\newtheorem{solution}{Solution}
\newtheorem{summary}{Summary}
\numberwithin{equation}{section}
\newenvironment{pf}[1][Proof]{\noindent{\it {#1.}} }{\ \rule{0.5em}{0.5em}}
\newenvironment{ex}[1][Example]{\noindent{\it {#1.}}}

\thispagestyle{empty}


\begin{center}

{\Large\scshape The mechanism why colliders could create quasi-stable black holes
\par}
\vskip15mm

\textsc{R. Torres}\footnote{E-mail: ramon.torres-herrera@upc.edu}
\par\bigskip
{\em
Department of Applied Physics, UPC, Barcelona, Spain.}\\[.1cm]
\vspace{5mm}

\textsc{F. Fayos}\footnote{E-mail: f.fayos@upc.edu}
\par\bigskip
{\em
Department of Applied Physics, UPC, Barcelona, Spain.}\\[.1cm]
\vspace{5mm}

\textsc{O. Lorente-Esp\'{i}n}\footnote{E-mail: oscar.lorente-espin@upc.edu}
\par\bigskip
{\em
Department of Physics and Nuclear Engineering, UPC, Barcelona, Spain.}\\[.1cm]
\vspace{5mm}

\end{center}

\begin{abstract}
It has been postulated that black holes could be created in particle collisions within the range of the available energies for nowadays colliders (LHC). In this paper we analyze the evaporation of a type of black holes that are candidates for this specific behaviour, namely, small black holes on a brane in a world with large extra-dimensions. We examine their evolution under the assumption that energy conservation is satisfied during the process and compare it with the standard evaporation approach. We claim that, rather than undergoing a quick total evaporation, black holes become quasi-stable. We comment on the (absence of) implications for safety of this result. We also discuss how the presence of black holes together with the correctness of the energy conservation approach might be experimentally verified.
\end{abstract}



\vskip10mm





\setcounter{equation}{0}


%
%
%
%
%

\section{Introduction}

It has been suggested that, if additional dimensions existed, then a brane could confine a subset or entire fields of the standard model. Gravitons, on the other hand, could propagate through the higher dimensional bulk and, as a result, gravity would be weak for observers living in the brane. In this way, the true scale of gravity, instead of being of the order of Planck's mass, could be as low as the electroweak scale, what would solve the \textit{hierarchy problem}. Moreover, black holes could be generated within the energy range available for nowadays colliders and, more specifically, for the particle collisions carried out at the Large Hadron Collider (LHC) \cite{ADD}\cite{AAD}.

Assuming that a black hole (BH) could be formed in this way, it would decay by emitting Hawking radiation. The evaporation process proceeds in some stages \cite{Gidd}. In the first stage, the \textit{balding phase}, the black hole radiates away its multi-pole moments inherited from the particles collision until it settles down in a hairless state. In the subsequent stage, the \textit{spin-down phase}, Hawking radiation carries away angular momentum until, finally, the non-rotating \textit{Schwarzschild phase} is reached. Then, the rest of the energy is emitted as Hawking radiation until the \textit{Planckian phase} is reached. Although a full quantum gravitational description would be required in this phase, it is usually assumed that the black hole should decay in some last few particles reaching a total evaporation. (Of course, this is not the only possibility and some authors have considered the formation of non-zero mass stable remnants. However, this alternative has received a lot of criticism \cite{Rem1}\cite{Rem2}\cite{Rem3}. Anyway, we will simply not consider stable remnants in this paper).


Let us recall that the pioneering approach for studying black hole evaporation \cite{Haw75} was based on results of quantum field theory on a fixed curved background (Schwarzschild's solution). Later, many different approaches have deduced the existence of Hawking evaporation from different points of view, but usually keeping the same assumption of a fixed background.
However, it is well-known that fixed background approaches
are not in agreement with energy conservation,
since the energy radiated by the black hole should be balanced by a corresponding decrease of its mass, what is unfeasible if the background has been fixed.
For macroscopic black holes their corresponding Hawking-Bekenstein temperatures are very small
($T_{BH}\approx 10^{-7} M_{Sun}/M_{BH}$) and one can make the plausible assumption that the process of evaporation is
quasistatic, being described very accurately by thermal radiation. Thus, for these black holes the use of a fixed background is justified.
However, as the last stages of black hole evaporation
are approached and the Hawking-Bekenstein temperature is not so small, the effects of energy conservation should become
non-negligible since the emission of a single particle could have a significant backreaction on the background.

Only recently \cite{P&W} Hawking radiation has been derived taking into account the back-reaction effect of the radiation
on the black hole thanks to the requirement of energy conservation.
Our aim in this paper is to follow this approach and apply it to the case of black holes with large additional compactified dimensions whose generation has been conjectured to occur within the range of energies in current colliders. We will try to analyze their evaporation and the relevance that energy conservation could have in the last stages of their evolution. In order to do this we will first recall the standard semiclassical calculations that lead to the usual result of black hole complete evaporation (what always includes to extend the calculations beyond their natural limit) \cite{Haw74}\cite{FN-S}. Then we will compare  these results with the application of the semiclassical calculations satisfying energy conservation (following the  approach in \cite{P&W}) to the whole evolution.
We will argue that
energy conservation might be the sufficient physical mechanism which could prevent the total evaporation of black holes created at the LHC.

Let us comment that our approach is not the only one to deal with BH evaporation under the requirement of energy conservation. Casadio and Harms \cite{Casad1,Casad2} have also demanded energy conservation by using a microcanonical description of black holes in which BHs are considered to be the excitation modes of p-branes.
Our approaches share some common features. However, our final results differ\footnote{In the specific (ADD) scenario treated in this paper they obtain total evaporation, but with slightly longer-lived BHs.} mainly due to what we believe is an oversimplification in the computations of the BH luminosity in \cite{Casad1,Casad2}.

The paper has been divided as follows. In section \ref{sectherm} we introduce the effective metric describing the black hole and we describe the BH properties from a strictly thermal point of view. This will allow us to model the thermal evaporation of the black hole and to describe its total evaporation. In section \ref{sectun} we study the black hole under the approach of energy conservation. This will lead us to a different effective distribution for the emitted photons as well as to a different luminosity which will later be used to model the evaporation process under the requirement of energy conservation. The subsequent BH evolution will be then contrasted with the thermal evolution found in section \ref{sectherm}. Finally, the results will be discussed in section \ref{seccon}. In particular, we will comment on the implications that quasi-stable black holes could have on the collider's safety and the chances of detecting the formation of these quasi-stable black holes.

\section{The thermal approach}\label{sectherm}

Since we are interested in the last stages of evaporation, we will assume that a black hole has been created in the collision of particles in a collider. Then, the black hole passes through the balding phase and the spin down phase arriving to the phase we are interested in: The Schwarzschild phase.

Let us first review the Schwarzschild phase for a black hole on a brane in a world with large extra-dimensions. We consider $d$ to be the dimension of the bulk spacetime and we assume that we live on a 3+1 dimensional brane. In this phase we assume that, if the black hole size is $R_0$, the extra dimensions will have size $L$ such that $R_0\ll L$. In this way, the geometry near the black hole is simply that of a d-dimensional Schwarzschild solution \cite{Park}
\begin{equation}
ds^2=-f(R) dt_S^2+\frac{dR^2}{f(R)}+R^2 d\Omega_{d-2}^2,
\end{equation}
where $\Omega_n$ denotes the surface of a unit n-sphere and
\begin{equation}
f(R)=1-\left(\frac{R_0}{R}\right)^{d-3},
\end{equation}
so that the event horizon ($f(R)=0$) is at $R=R_0$.

The induced metric on the brane for $R\sim R_0$ (where all the interesting physics takes place) will be
\begin{equation}\label{fR}
ds^2=- f(R) dt_S^2+\frac{dR^2}{f(R)}+ R^2 d\Omega_2^2.
\end{equation}

The event horizon $R=R_0$ has a corresponding surface gravity
\begin{equation}\label{surfgrav}
\kappa= \frac{1}{2} \left.\frac{df}{dR}\right\rfloor_{R=R_0}=\frac{d-3}{2 R_0}.
\end{equation}
As result, for instance, of a standard Euclidean continuation of the fixed static geometry through $R_0$, one gets that the outer horizon should emit Hawking radiation with a thermal distribution of temperature
\begin{equation}\label{temp}
T=\frac{\kappa}{2 \pi}=\frac{d-3}{4 \pi R_0}.
\end{equation}

As we stated, (\ref{fR}) is only valid approximation for $R\sim R_0 \ll L$. However, for $R\gg L$ one expects to recover the standard (3+1) Schwarzschild solution in which
\[
f(R)=1-\frac{2 G_4 M}{R},
\]
where $G_4$ is the usual (four dimensional) Newton's gravitational constant and $M$ is the mass of the black hole. From here it can be deduced \cite{Empa} that the mass $M$ and the black hole size $R_0$ are related through
\begin{equation}\label{R0}
R_0=
\beta_d\, M^{1/(d-3)},
\end{equation}
where
\[
\beta_d\equiv\left(\frac{16 \pi G_d}{(d-2) \Omega_{d-2}} \right)^\frac{1}{d-3}.
\]
and $G_d=G_4 L^{d-4}$ is the $d$-dimensional Newton's constant.


The existence of a temperature associated with the event horizon implies a standard thermal distribution for the emitted photons
\begin{equation}\label{nStand}
<n(E)>_{Stand.}=\frac{1}{\exp(E/T)-1},
\end{equation}
where $E$ is the photon energy. These photons will be mainly radiated on the brane \cite{Empa}\cite{J&P} and the total flux of radiated energy \cite{FN-S} measured from an observer in the asymptotically flat region of the brane will be
approximately given by
\begin{equation}\label{lstand}
L_{Stand.} \simeq \frac{1}{2 \pi}\int_0^\infty <n(E)>_{Stand.} \gamma_{Stand.}(E) E dE,
\end{equation}
where $\gamma_{Stand.}(E)$ is the greybody factor that takes into account the amount of photons that will be emitted by the BH and will be later backscattered by the effective potential resulting from the spacetime geometry.


\subsection{Backreaction: A toy model of BH evaporation}\label{ssBR}


In order to get an idea of the total evaporation of black holes in this thermal approach, it will be enough to consider the back-reaction due to the emission of radiation.
Let us first write the effective metric (\ref{fR}) in terms of ingoing Eddington-Finkelstein-like coordinates $\{u,R,\theta,\varphi\}$, where
\[
u=t_S+\int^R \frac{dR'}{f(R)}\ ,
\]
as
\begin{equation}\label{ScEF}
ds^2=- f(R) du^2+2 du dR + R^2 d\Omega^2.
\end{equation}
Now, we can model the mass lost taking into account the heuristic picture that describes Hawking radiation as due to a tunneling process. I.e., whenever a pair of virtual particles is created, when the particle with positive energy escapes to infinity its companion, with negative energy, falls into the black hole thus reducing the BH mass. In this way, if we consider negative energy massless particles following ingoing null geodesics $u=$constant, the mass at infinity of the black hole becomes a decreasing function $M(u)$. The metric which incorporates the effect of the decreasing BH mass due to the ingoing null radiation
is (\ref{ScEF}) with $M$ replaced by $M(u)$
\begin{equation}\label{Vaid}
ds^2=- f(R; M(u)) du^2+2 du dR + R^2 d\Omega^2,
\end{equation}
where we have now made explicit the new dependence of $f$ on $u$ through the evolving black hole mass: $f(R; M(u))$.
On the other hand, the flux of negative energy particles directed towards the black hole equals the flux of outgoing radiated particles that reach the future lightlike infinity and, therefore,
\begin{equation}\label{evoleq}
\frac{dM}{du}=-L_{Stand.}(M)
\end{equation}

The solution of this differential equation provide us with the approximate evolution of the black hole mass and the expected evaporation. While the total evaporation time that one obtains depends on the specific form of the greybody factor in (\ref{lstand}), it is usually assumed to happen in an extremely short time from around $10^{-27}$ to $10^{-25}$ seconds \cite{Empc}.


\section{The energy conservation approach}\label{sectun}

Let us now consider Hawking radiation coming out from the black hole
thanks to the tunneling process occurring through the event horizon $R_0$ and taking into account the consequences of energy conservation. In order to do this, we will follow the steps described in \cite{P&W} for the standard (3+1) Schwarzschild case by adapting them to the case under consideration. First, we rewrite the induced metric (\ref{fR}) in Painlev\'e-like coordinates \cite{Pain} so as to have coordinates which are not singular at the horizon. It suffices to introduce a new coordinate $t$ replacing the Schwarzschild-like time $t_S$ such that $t=t_S+h(R)$ and fix h(R) by demanding the constant time slices to be flat. In this way one gets:
\begin{equation}
ds^2=- f(R) dt^2+2 \sqrt{1-f(R)} dt dR+ dR^2 + R^2 d\Omega^2,
\end{equation}
In these coordinates the radial null geodesics describing the evolution of \emph{test} massless particles are given by
\begin{equation}\label{geodtest}
\frac{dR}{dt}=\pm 1-\sqrt{1-f(R)}
\end{equation}
with the upper (lower) sign corresponding to outgoing (ingoing, respectively) geodesics.


In \cite{K&W1}\cite{K&W2}\cite{P&W}\cite{I&Y} it was found that, when a self-gravitating shell of energy $E$ travels in
a spacetime characterized by an ADM mass $M$,
the geometry outside the shell is also characterized by $M$, but energy conservation implies that
the geometry inside the shell is characterized by $M-E$. It was also found that the shell
then moves on the geodesics given by the interior line element.
In this way, according to (\ref{geodtest}), one expects a shell of energy $E$ to satisfy the evolution equation
\begin{equation}\label{geodshell}
\frac{dR}{dt}=\pm 1-\sqrt{1-f(R;M-E)}.
\end{equation}
where $f(R;M-E)$ is $f(R)$ with the mass $M$ replaced by $M-E$.

%

Let us now consider pair production occurring just beneath the event horizon with a positive energy particle tunneling out.
The standard results of the WKB method for the tunneling through a potential barrier that would be classically forbidden can be directly applied due to the infinite redshift near the horizon \cite{P&W}. In particular, the semiclassical emission rate will be given by $\Gamma \sim \exp\{-2 \mbox{Im} S\}$, where $S$ is the particle action. Therefore, we have to compute the imaginary part of the action for an
outgoing positive energy particle which crosses the horizon $R_0$ outwards from $R_{in}$ to $R_{out}$.
\begin{equation}
\mbox{Im} S=\mbox{Im} \int_{R_{in}}^{R_{out}} p_R dR= \mbox{Im}  \int_{R_{in}}^{R_{out}}  \int_{0}^{p_R} dp'_R dR,
\end{equation}
where we have taken into account that only $p_R$, the $R$-component of the four-momentum,  contributes to the imaginary part of the action. Using Hamilton's equation $\dot{R}=+dH/dp_R\rfloor_R$ and $H=M-E'$, this can be written with the help of (\ref{geodshell}) as
\begin{eqnarray}\label{imS}
\mbox{Im} S&=& \mbox{Im} \int_{M}^{M-E} \int_{R_{in}}^{R_{out}} \frac{dR}{\dot R}  dH=\nonumber\\
&=&\mbox{Im} \int_{0}^{E} \int_{R_{in}}^{R_{out}} \frac{dR}{1-\sqrt{1-f(R;M-E')}} (-dE').
\end{eqnarray}

Then by deforming the contour of integration so as to ensure that positive energy solutions decay in time, considering that a particle just inside the horizon tunnels just outside a shrunken horizon ($R_{in}>R_{out}$) and
taking into account that close to the horizon
\begin{eqnarray*}
f(R; M-E') &\simeq& \left.\frac{\partial f(R;M-E')}{\partial R}\right\rfloor_{R=R_0(M-E')} (R-R_0(M-E'))\\
&=&\frac{d-3}{R_0(M-E')} (R-R_0(M-E')),
\end{eqnarray*}
where $R_0(M-E')$ is the position of the outer horizon when the BH mass at infinity is $M-E'$,
one gets
\[
 \int_{R_{in}}^{R_{out}} \frac{dR}{1-\sqrt{1-f(R;M-E')}} =-i 2 \pi \frac{R_0(M-E')}{d-3}.
\]

We can then write (\ref{imS}) as
\begin{equation}\label{partchan}
\mbox{Im} S= \int_{0}^{E}  \frac{2 \pi R_0(M-E')}{d-3} dE' = \frac{2 \pi \beta_d}{d-2} \left[M^\frac{d-2}{d-3}-(M-E)^\frac{d-2}{d-3}  \right].
\end{equation}

Tunneling also happens when a pair is created outside the horizon and the negative energy particle tunnels into the black hole.
Then, following the procedure for the Schwarzschild case in \cite{P&W}, the imaginary part of the action for this ingoing particle satisfies
\begin{equation}
\mbox{Im} \int_{0}^{-E} \int_{R_{out}}^{R_{in}} \frac{dR}{-1+\sqrt{1-f(R;M-E')}} dE'=\int_{0}^{E}  \frac{2 \pi R_0(M-E')}{d-3} dE',
\end{equation}
what coincides with the expression for the previous channel (\ref{partchan}).
Both channels contribute to the rate of the Hawking process, but we have seen that both contributions provide us with the same exponential term for the semiclassical rate
\begin{equation}\label{emprob}
\Gamma\sim e^{-2 \mbox{\scriptsize Im} S }=\exp\left(- 4 \pi \frac{\beta_d}{d-2} \left[M^\frac{d-2}{d-3}-(M-E)^\frac{d-2}{d-3}  \right]\right).
\end{equation}

If quadratic terms in $E$ could be neglected
we could develop Im $S$ up to first order in $E$ as
\[
\mbox{Im} S \simeq \frac{2 \pi \beta_d}{d-3} M^\frac{1}{d-3} E=\frac{2 \pi }{d-3} R_0 E,
\]
obtaining a thermal radiation for the black hole ($\Gamma \sim \exp\{-E/T\}$) with temperature
\begin{equation}\label{TQBH}
T=
\frac{d-3}{4 \pi R_0},
\end{equation}
that coincides with the expected temperature (\ref{temp}).


%


\subsection{Spectral distribution and flux}\label{seclum}
Notwithstanding the comments about the temperature of the black hole, it is important to remark that the higher order terms in $E$, neglected in (\ref{TQBH}), imply a deviation from pure thermal emission.
If we consider the full consequences of energy conservation, now the distribution function for the emission of photons
is not the standard Boltzmann distribution (\ref{nStand}), but the distribution (see \cite{K&K} --correcting the result in \cite{K&W1,K&W2}--)
\begin{displaymath}
<n(E)>_{EC}=\frac{1}{\exp \left(2 \mbox{Im} S \right)-1}.
\end{displaymath}
For our induced metric this can be written as
\begin{equation}\label{nE}
<n(E)>_{EC}=\frac{1}{\exp \left(4 \pi \frac{\beta_d}{d-2} \left[M^\frac{d-2}{d-3}-(M-E)^\frac{d-2}{d-3}  \right] \right)
-1},
\end{equation}
where $0< E \leq M$ (i.e., the black hole cannot emit more energy than its own mass).

Now, we would like to use this distribution in order to write the flux of radiation when energy conservation is taken into account. This requires the use of an appropriate greybody factor.
Since photons are mainly radiated on the brane, it can be shown \cite{K&MR} that
(without taking into account energy conservation)
for any static spherically symmetric black hole with outer horizon $R_0$, and whenever $E R_0\ll 1$, the greybody factor corresponding to a wave with total angular momentum number $j$ ($j=1,2,\ldots$) takes the form
\begin{equation}\label{gammast}
\gamma(E;M;j)=4 E^2 R_0(M)^2 (2 E R_0(M))^{2 j} \frac{2 j +1}{(d-3)^2} \left( \frac{\Gamma(\frac{j}{d-3})\Gamma(\frac{j+1}{d-3})\Gamma(j+2)}{\Gamma(\frac{2j+1}{d-3})\Gamma(2j+2)} \right)^2.
\end{equation}
Note that now we have also emphasized the dependence on $M$ through $R_0$ and that the previously defined greybody factor in (\ref{lstand}) is obtained using $\gamma_{Stand.}(E)=\sum_j \gamma(E;M;j)$.
The expression (\ref{gammast}) is an excellent approximation for the case we are interested in ($M\sim 0$) since, in the energy conservation approach, \emph{all} photons emitted by the black hole will satisfy $E R_0\ll 1$ for small enough black hole masses\footnote{Note that this is a very stringent test since, in order for the greybody factor to be a good approximation, it is not necessary for \textit{all} photons to satisfy the inequality. It would be enough to demand to their average energy $<E>$ to satisfy it. Moreover, in \cite{K&MR} it is shown that it would be enough to require $<E> R_0< 0.4$.}. In effect, since the maximum possible energy for an emitted photon is $E=M$, if all photons satisfied the inequality then we would have to demand $M R_0\ll 1$. From (\ref{R0}), this will happen in the last stages of evaporation when the black hole mass will have to satisfy
\[
M\ll \left( \frac{(d-2) \Omega_{d-2}}{16 \pi G_d} \right)^\frac{1}{d-2}.
\]
For example, if one considers the generalized Planck scale in $d$ dimensions \cite{R&S},
\[
M_d = (8 \pi G_d)^\frac{-1}{d-2},
\]
if $d=7$ the approximation will be excellent for $M\ll 2.4 M_7$, while if $d=11$ the approximation will be excellent for $M\ll 1.7 M_{11}$.

If we want to take into account how the back-reaction due to the emission of a photon with energy $E$ affects the greybody factor, one has to consider that, if an outgoing shell is
back-scattered, thus becoming ingoing, the shell must have always moved on null geodesics given by a line element with
mass $M-E$ \cite{RPO}. In this way, the greybody factor including the back-reaction ($\gamma_{EC}$) corresponding to a wave with total angular momentum $j$ can be found directly using (\ref{gammast}) as
\begin{equation}\label{gammaEC}
\gamma_{EC}(E;M;j)=\gamma(E;M-E;j).
\end{equation}

\begin{figure}
\includegraphics[scale=1]{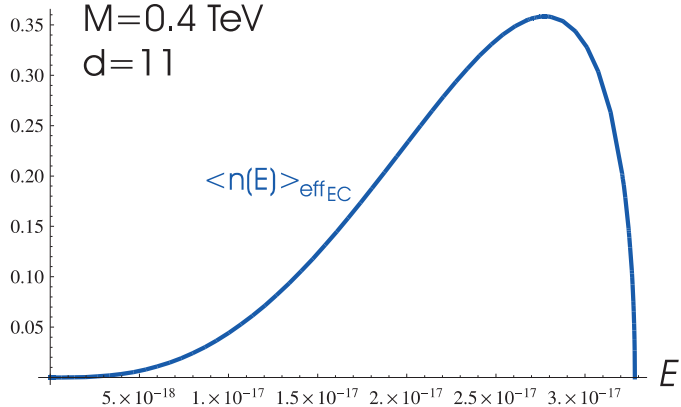}
\caption{\label{figdist} Effective distribution in the energy conservation approach for a black hole of mass $M=0.4$ TeV. The dimension of the bulk has been chosen to be $d=11$. The magnitude $L$ of the extra compact dimensions has been chosen so that it can provide a solution to the hierarchy problem, i.e. $M_d\sim 1$ TeV.
Note that, contrarily to the thermal approach, the distribution is only defined in the range $0<E\leq M$ (Planck's units in the figure).
}
\end{figure}

In order to compare the effective distribution (i.e., the distribution of photons as seen from an observer far away) in the thermal and the energy conservation approach, one has to compare
\[
<n(E)>_{eff_{Stand.}}\equiv <n(E)>_{Stand.} \gamma_{Stand.}
\]
with
\[
<n(E)>_{eff_{EC}}\equiv <n(E)>_{EC} \gamma_{EC}.
\]
It can be easily shown that, if $L$ is chosen such that it solves the \textit{hierarchy problem} ($M_d \sim 1 \mbox{TeV}$) \cite{Park}, these effective distributions are practically identical for black holes with masses
much bigger than $M_d$.
However, for masses of the order of (or order smaller) than $M_d$
the effective distributions differ significantly.
On the one hand, the effective distribution in the thermal approach provides much bigger values in all the range of energies and, on the other hand, the thermal approach allows the emission of a non-negligible amount of high-energy photons, what is not allowed in the energy conservation approach. In fact, the effective distribution in the energy conservation approach is only defined for $0<E\leq M$, what implies that only long wavelength photons will be emitted for $M\ll M_d$ (see an example in figure \ref{figdist}).
We can now use (\ref{nE}) and the greybody factor $\gamma_{EC}$ in order to write the flux of radiation when energy conservation is satisfied as
\begin{equation}
L_{EC}\simeq \frac{1}{2 \pi} \sum_j \int_0^M <n(E)>_{EC}  \gamma_{EC}(E;M;j) E dE.
\label{lumi}
\end{equation}
Note that the fact that the range of energies for the emitted particles has to be $0< E \leq M$ is reflected in the integration limits.

\begin{figure}
\includegraphics[scale=1.0]{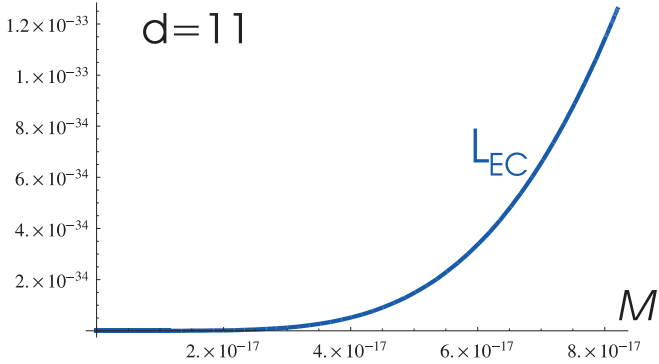}
\caption{\label{luminosidades_r} Luminosity of the black hole as a function of its mass $M$ (in Planck's units) for an 11-dimensional bulk.
The magnitude of the compact extra-dimensions has been chosen so that it provides a solution to the hierarchy problem: for $d=11$, $L=2\cdot 10^{-14}$ m.
The luminosities plotted correspond to a range of masses $0\leq M \leq 1 TeV$.
Note that, in the energy conservation approach, the luminosity vanishes as the black hole mass tends to zero.}
\end{figure}

It is important to remark that, as expected, for masses much bigger than the generalized Planck scale in $d$ dimensions
the luminosity taking into account energy conservation (\ref{lumi}) is nearly identical to the luminosity in the thermal approach.
Only for masses bellow the generalized Planck scale the differences between luminosities are considerable. The extreme case happens for masses close to zero since the thermal approach provide us with unbound luminosities, while the energy conservation approach provide us with luminosities that are close to zero (see figure \ref{luminosidades_r}).

\subsection{Evaporating BH model in the energy conservation approach}
In order to model the evaporation of the black hole under the energy conservation approach we just have to re-consider the steps followed in subsection \ref{ssBR} for the thermal case. In this way, the mass evolution when energy conservation is satisfied trivially lead us to the differential equation
\begin{equation}\label{evolEC}
\frac{dM}{du}=-L_{EC}.
\end{equation}

In particular, in the last stages of evaporation the expected behaviour can be computed taking into account that
for $M\sim 0$ we have $E R_0\sim 0$ and we see that the main contribution to the greybody factor (\ref{gammaEC}) (from (\ref{gammast})) is due to the $j=1$ term which could be written as
\begin{equation}\label{gammaused}
\gamma_{EC}(E;M)= C_d E^4 R_0^4(M-E),
\end{equation}
where
\[
C_d\equiv \frac{48}{9 (d-3)^2} \left( \frac{\Gamma(\frac{1}{d-3})\Gamma(\frac{2}{d-3})}{\Gamma(\frac{3}{d-3})} \right)^2.
\]
In this way, using this in (\ref{lumi}) the luminosity for $M\sim 0$ will be approximately given by
\[
L_{EC}\simeq \frac{(d-2) C_d \beta_d^3  }{8 \pi^2}\int_0^M \frac{E^5 (M-E)^\frac{4}{d-3}}{M^\frac{d-2}{d-3}-(M-E)^\frac{d-2}{d-3}} dE
\]
what provide us with
\begin{equation}\label{LECM0}
L_{EC}\simeq \frac{d-2}{8 \pi^2} C_d \beta_d^3 I_d\ M^\frac{5 d-12}{d-3},
\end{equation}
where $I_d$ is the finite positive result of the integral
\[
I_d\equiv \int_0^1 \frac{x^5 (1-x)^\frac{4}{d-3}}{1-(1-x)^\frac{d-2}{d-3}} dx.
\]
If we solve the evolution equation (\ref{evolEC}) for $M\sim 0$ using (\ref{LECM0}) we get
\[
M(u)=\left[M_0^{-\frac{4 d-9}{d-3}}+ \frac{(d-2)(4 d-9)}{8 \pi^2 (d-3)} C_d \beta_d^3 I_d\  u \right]^{-\frac{d-3}{4 d-9}}.
\]
For $u$ big enough
\[
M(u)\sim u^{-\frac{d-3}{4 d-9}},
\]
what, taking into account that $d> 4$, only asymptotically approaches zero for $u\rightarrow \infty$.

\begin{figure}
\includegraphics[scale=.8]{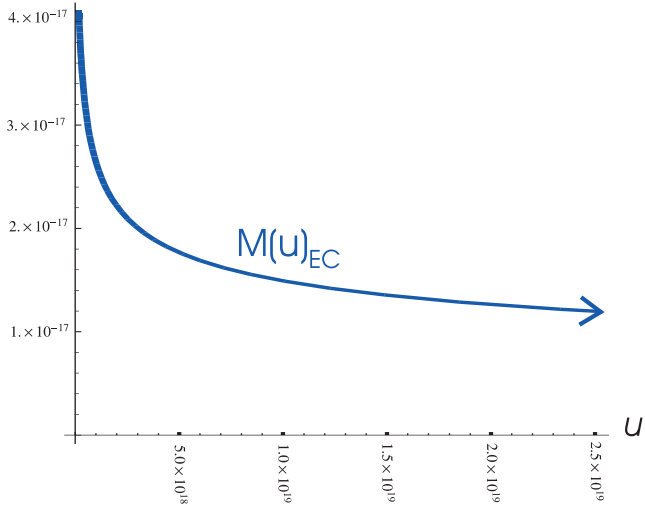}
\caption{\label{figevol} The evolution of a black hole in a $d=11$ bulk from $M=0.5$ TeV to $M=0.1$ TeV.
After the mass $M=M_d \sim 1$ TeV has been reached the evolution slows down drastically. The apparent initial steep slope for $M\simeq 0.5$ TeV is only due to the fact that the evolution is even slower as time passes. The mean lifetime of the black hole becomes infinite. }
\end{figure}

The evolution of a black hole could then be summarized as follows. Since for $M\gg M_d$, the luminosities are very similar in the thermal and the energy conservation approach, the predicted evolutions will also be similar until the mass reaches values of the order $M\sim M_d$. Then, the evolution will slow down according to the energy conservation approach in such a way that the mean lifetime of the BH will be infinite.
(Note that, on the contrary, the thermal approach usually assumes a speed up of the evolution that would have ended in a final BH evaporation).
The specific evolution from 0.5 TeV to 0.1 TeV in the energy conservation approach (i.e., by computing the numerical solution of (\ref{evolEC})) is shown in figure \ref{figevol}.

\section{Conclusions}\label{seccon}
In this paper we have
studied the evolution of black holes on a brane in a world with large extra-dimensions
when energy conservation is satisfied. In order to do this we have performed a simple model of black hole evaporation in which radiation is emitted mainly from the black hole into the brane (what allows as to use the induced metric (\ref{fR})). While in this model the black hole completely evaporates when one follows the thermal approach, if energy conservation is enforced, the black hole first evaporates in a rate similar to that of the thermal approach (for $M\gg M_d$) and then, it slows down in its late stages (for $M\lesssim M_d$). In this way, it never reaches zero mass. Since the evaporation is really a probabilistic process in which discrete amounts of energy (quanta) are emitted by the black hole, the exhibited smooth evolution in the last stages should be seen as the mean evolution for black holes which, by means of energy conservation, become \textit{quasi-stable}.
%
%

Of course, we do not argue that the semiclassical calculations carried in this paper represent a faithful representation of the last stages of black hole evaporation since this could only be accomplished by using a full Quantum Gravity Theory. What we suggest is that energy conservation might be a sufficient condition for avoiding the total evaporation of black holes that could be produced in colliders. The physical mechanism behind this would be that, contrarily to the assumption in the thermal approach, energy conservation implies that a black hole cannot emit particles of arbitrary high energy.
Rather, the picture offered by the energy conservation approach suggests that in the last stages of the evaporation only long wavelength particles could tunnel out the horizon, what eventually would prevent the total evaporation.

It has been conjectured that, if a mechanism existed such that black holes could live long enough to escape the LHC and penetrate into the Earth, they could grow due to the accretion of matter, what would eventually lead to a catastrophic scenario.
It is important to remark that this conjecture depends on the efficiency of the two accretion mechanisms. Namely, on the one hand, the black hole could use its gravitational force on surrounding matter in order to swallow it (a mechanism known as Bondi accretion).
On the other hand, the black hole could accrete matter by colliding with atomic or sub-atomic particles in its way through the Earth.
However, it has been shown \cite{CFH} that accretion through both mechanisms would be only appreciable for black holes with horizons much bigger than the ones considered here (whose horizons in their quasi-stable phase satisfy $0<R_0\lesssim 10^{-19}$ m $\ll L$). Therefore, we arrive to the expected result\footnote{That black holes formed through particle collisions can not be dangerous is well-known due to the existence of stable astronomical bodies that are continuously exposed to high-energy cosmic-ray collisions \cite{G&M}.} that the long lived black holes predicted by the energy conservation approach can not be dangerous.

It has been argued that the LHC could produce about one black hole per second and that
the production of photons would be noticeably increased if BH do form, what could be used for their detection \cite{D&L}. Moreover, according to the thermal picture black holes have to decay into hard photons (as well as high energy jets and leptons). Consequently, experiments to discover black holes are designed taking this picture into account \cite{CMS1}\cite{CMS2}. However, according to the energy conservation approach the energy of the emitted photons is bounded
and the emission will consist mainly of soft photons for  $M\ll M_d$.
The prediction by the energy conservation approach of the absence of final decay higher-energy particles and the later emission of only long wavelength photons in the latest stages of the BH evolution is a clear signature that differentiates it from the thermal picture.
Furthermore, the presence of a quasi-stable black hole should provide some missing matter in the final products of the collision plus evaporation of the order of $M\lesssim M_d$ after the thermal complete evaporation time has been reached ($\lesssim 10^{-25}$ s). In fact, after $0.1$ s has passed from the formation of the black hole
the mean remaining mass should still be of the order of $1$ MeV.

In summary, if black holes could be produced at the LHC at around 10 TeV, then an increase in the productions of photons could be used to detect their presence. However, if the energy conservation picture in the scenario of large extra-dimensions is correct, only about a 90\% of the black hole initial mass should be emitted as high-energy particles in about $10^{-25}$ s. Then, the higher-energy particles (that should have been emitted according to the thermal approach) should be missed. The remaining 10\% of the mass should then be emitted in the form of, mainly, relatively soft particles which should be more easily detectable in the first second after the generation of the black hole.





\begin{thebibliography}{99}
\bibitem{ADD}
Arkani-Hamed N, Dimopoulos S and Dvali G R 1998 {\it Phys. Lett. B} {\bf 429} 263
  (arXiv:hep-ph/9803315)
\bibitem{AAD}
Antoniadis I, Arkani-Hamed N, Dimopoulos S and Dvali G R 1998 {\it Phys. Lett. B} {\bf 436} 257 (arXiv:hep-ph/9804398)
\bibitem{Gidd}
Giddings S B and Thomas S 2002 {\it Phys. Rev. D} {\bf 65} 056010 (arXiv:hep-ph/0106219)
\bibitem{Rem1}
Preskill J 1992 {\it Proceedings of the International Symposium on Black Holes, Membranes, Wormholes and Superstrings} (arXiv:hep-th/9209058)
\bibitem{Rem2}
Giddings S B 1994 {\it Phys. Rev. D} {\bf 49} 947
\bibitem{Rem3}
Susskind L 1995 (arxiv:hep-th/9501106)
\bibitem{Haw75}
Hawking S W 1975 {\it Commun. Math. Phys.} {\bf 43} 199
\bibitem{P&W}
Parikh M K and Wilczek F 2000 {\it Phys. Rev. Lett.} {\bf 85} 5042 (arXiv:hep-th/9907001v3)
\bibitem{Haw74}
Hawking S W 1974 {\it Nature} {\bf 248} 30
\bibitem{FN-S}
Fabbri A and Navarro-Salas J 2005 {\it Modeling Black Hole Evaporation} London: Imperial College Press
\bibitem{Casad1}
Casadio R and Harms B 2011 {\it Entropy} {\bf 13} 502
\bibitem{Casad2}
Casadio R and Harms B 2002 {\it Int. J. Mod. Phys. A} {\bf 17} 4635
\bibitem{Park}
Park S C 2012 {\it Prog. Part. Nucl. Phys.} {\bf 67} 617 (arXiv:hep-ph/1203.4683)
\bibitem{Empa}
Emparan R, Horowitz G T and Myers R C 2000 {\it Phys. Rev. Lett.} {\bf 85} 499 (arXiv:hep-th/0003118)
\bibitem{J&P}
Jung E and Park D K 2007 {\it Nucl. Phys. B} {\bf 766} 269 (arXiv:hep-th/0610089)
\bibitem{Empc}
Emparan R 2003  (arXiv:hep-ph/0302226)
\bibitem{Pain}
Painlev\'e P 1921 {\it C. R. Acad. Sci. (Paris)} {\bf 173} 677
\bibitem{K&W1}
Kraus P and Wilczeck F 1995 {\it Nucl. Phys.} {\bf B433} 403 (arXiv:hep-th/9408003)
\bibitem{K&W2}
Kraus P and Wilczeck F 1995 {\it Nucl. Phys.} {\bf B437} 231 (arXiv:hep-th/9411219)
\bibitem{I&Y}
Israel W and Yun Z 2010 {\it Phys. Rev. D} {\bf 82} 124036
\bibitem{K&K}
Keski-Vakkuri E and Kraus P 1997 {\it Nucl. Phys.} {\bf B491} 249 (arXiv:hep-th/9610045)
\bibitem{K&MR}
Kanti P and March-Russell J 2003 {\it Phys.Rev. D} {\bf 67} 104019
\bibitem{R&S}
Randall L and Sundrum R 1999 {\it Phys. Rev. Lett.} {\bf 83} 3370
\bibitem{RPO}
Torres R, Fayos F and Lorente-Esp\'{i}n O 2013 {\it Phys. Lett. B} {\bf 720} 198
\bibitem{CFH}
Casadio R, Fabi S and Harms B 2009 {\it Phys. Rev. D} {\bf 80} 084036 (arXiv:hep-ph/0901.2948)
\bibitem{G&M}
Giddings S B and Mangano M L 2008 {\it Phys. Rev. D} {\bf 78} 035009
  (arXiv:hep-ph/0806.3381)
\bibitem{D&L}
Dimopoulos S and Landsberg G 2001 {\it Phys. Rev. Lett.} {\bf 87} 161602 (arXiv:hep-ph/0106295)
\bibitem{CMS1}
CMS Collaboration 2010 CMS-EXO-10-017, CERN-PH-EP-2010-073 (arXiv:hep-ex/1012.3375)
\bibitem{CMS2}
CMS Collaboration 2013 CMS-EXO-12-009, CERN-PH-EP-2013-043 (arXiv:hep-ex/1303.5338)
\end{thebibliography}
\end{document}